\documentclass{aa}
\usepackage[centertags]{amsmath}
\usepackage{epsfig,amssymb,xspace}

\begin{document}

\thesaurus{}

\title{Microlensing Optical Depth of the Large Magellanic Cloud}
\author{
{\'E}.~Aubourg\inst{1},
N.~Palanque-Delabrouille\inst{1},
P.~Salati\inst{2,3},
M.~Spiro\inst{1},
R.~Taillet\inst{2,3}
}
\institute{
CEA, DSM, DAPNIA,
Centre d'{\'E}tudes de Saclay, 91191 Gif-sur-Yvette Cedex, France
\and LAPTH, chemin de Bellevue, BP 110, 74941 Annecy-le-Vieux Cedex, France.
\and Universit{\'e} de Savoie, B.P. 1104, 73011 Chamb{\'e}ry Cedex, France.
}
\offprints{Nathalie.Delabrouille@cea.fr}

\date{Received;accepted}

\authorrunning{Aubourg et al.}
\titlerunning{Microlensing Optical Depth of the {\sc lmc}}

\maketitle

\def\etal{{et al.\xspace}}
\def\eros{{\sc eros}\xspace}
\def\macho{{\sc macho}\xspace}
\def\lmc{{\sc lmc}\xspace}
\def\smc{{\sc smc}\xspace}
\def\ms{{\sc ms}\xspace}
\def\ie{{\em i.e.}\xspace}
\def\Msol{M_\odot}
\def\Lsol{L_\odot}
\def\tauLMC{\tau_{\mathrm{\mathsc{lmc}}}}
\def\DLMC{D_{\mathrm{\mathsc{lmc}}}}

\begin{abstract}
The observed microlensing events towards the \lmc do not have yet a 
coherent explanation.  If they are due to Galactic halo objects, the 
nature of these objects is puzzling --- half the halo in dark 
$0.5\;\Msol$ objects.  On the other hand, traditional models of the 
\lmc predict a self-lensing optical depth about an order of magnitude 
too low, although characteristics of some of the observed events favor 
a self-lensing explanation.  We present here two models of the \lmc 
taking into account the correlation between the mass of the stars and 
their velocity dispersion: a thin Mestel disk, and an ellipsoidal 
model.  Both yield optical depths, event rates, and event duration 
distributions compatible with the observations. The grounds for
such models are discussed, as well as their observational consequences.

\keywords: {Galaxy: halo, kinematics and dynamics, stellar content --
Cosmology: dark matter, gravitational lensing}
\end{abstract}

\section{Introduction}
\label{sec:Introduction}
One of the greatest uncertainties in the interpretation of the 
microlensing events detected towards the Large and the Small 
Magellanic Clouds by the \macho 
(Alcock et al. 1997a, \cite{MachoSMC}) 
and the \eros experiments 
(Renault et al. 1997, Palanque-Delabrouille et al. 1998) 
resides in the determination of the location and consequently of the 
nature of the lenses.

The \macho results can be converted into an optical depth of 
$2.9_{-0.9}^{+1.4} \times 10^{-7}$, while the two candidate events of 
\eros would account for $\sim 0.8 \times 10^{-7}$.  Combining both 
results leads to an average optical depth of $2.1_{-0.8}^{+1.3} \times 
10^{-7}$ (\cite{BennettReport}).

While this observed optical depth towards the 
\lmc is too small to account for a full dark halo of compact objects
(expected $\tau = 5 \times 10^{-7}$), it 
is much too large to be explained by the contribution from known
populations of low-mass stars in the disks of the Milky Way and the
\lmc: based on star counts, the contribution from the disk of our
Galaxy is found to be $\tau_{\rm MW} \sim 8 \times 10^{-9}$
(\cite{Gould97}), and the global rate from known stars in both
galaxies has been estimated by several authors to be $\lesssim20\%$
(\cite{wu94}, \cite{sahu94}, \cite{derujula95}) of that
observed.

Moreover, the location of a few lenses has been determined, either 
through the non-detection of parallax effect for 97-SMC-1 
(\cite{ErosSMC1}), or through the observation of the caustic crossing 
in the case of binary lenses, for 98-SMC-1 (Afonso et al. 1998, 
\cite{PlanCaus}, \cite{MachoCaus}, 
Rhie et al. 1999, Udalski et al. 1999) or 
MACHO-LMC-9  (Bennett et al. 1996).
The case for the latter is still unclear
however, since the measured lens velocity, incompatible with a
Galactic halo lens unless
the source happens to be a binary star with the appropriate
separation, seems too low for a \lmc disk star.

It has been pointed out by several authors 
(Kerins \& Evans 1999, 
\cite{DiSt99}) that the fact that the observed binary lenses are all 
located in the Clouds might indicate that {\em all} observed events 
arise from Clouds deflectors.  Indeed, if the two observed binary 
lenses with caustic crossings are in the Clouds, then the expected 
number of self-lensing caustic crossing events has to be greater than 
.53 at 90\% C.L. A very conservative estimate of the fraction of
caustic crossing events (\cite{MaPa91}, \cite{DiSt99})
implies that we expect more than 5 background self-lensing events, 
which is far from negligible as compared to the total number of
observed events.

The case for self-lensing in the \lmc and \smc is thus still open, 
despite the ``proof'' (\cite{Gould95}) that its contribution to the 
optical depth is too low.  As shown in (Palanque-Delabrouille et al 
1998), the \smc could be thick enough to yield an expected optical 
depth as large as $\sim 2 \times 10^{-7}$, compatible with the current 
observations towards this line-of-sight.  A more thorough exploration 
of possible \lmc models therefore appears worthwhile.

Gould (1995) noticed that the optical depth associated to the
self-lensing of a stellar disk, a good approximation for the bulk of 
the \lmc matter, depends only on the vertical velocity dispersion
$\sigma_{W}$ of the stars:
\begin{equation}
\tauLMC \; = \; 2 \, \frac{\sigma_{W}^{2}}{c^{2}} \, {\rm sec}^{2} i \;\; ,
\label{GOULD}
\end{equation}
where $i$ is the inclination of the disk.
Measurements of stellar velocities in the \lmc are difficult because
of its distance. Only the brightest stars are resolved. Gould based his
analysis on the CH objects studied by Cowley \& Hartwick (1991). These carbon
stars belong to the asymptotic giant branch (AGB) of the HR diagram. With an
apparent magnitude $m_{\rm bol} \sim 14-15$ and assuming a distance
modulus of 18.6, their bolometric luminosities are in the range
$2-5 \times 10^{3} \; \Lsol$. A velocity dispersion of $\sim$ 20 km/s 
led Gould to the conclusion that the optical depth for \lmc
self-lensing could not exceed $1.1 \times 10^{-8}$, well below the measured
value of $1-3 \times 10^{-7}$. However, carbon stars are a peculiar 
population. They have left the main sequence (\ms) on which they have spent
at most $\sim 10-12$ Gyr, the age of the \lmc. The average lifetime
of a star on the \ms decreases with the initial stellar mass $m$ as
\begin{equation}
t_{\rm \mathsc{ms}} \simeq
\left( 1.2 \times 10^{10} \; {\rm yr} \right) \,
\left( m \; [\Msol] \right)^{-3} \;\;,
\end{equation}
so that the oldest carbon stars observed today had an initial mass
larger than $\sim$ 1 $\Msol$. Assuming a Salpeter {\sc imf} for the \lmc
(Geha et al. 1998) and adopting the star formation history derived
by the same authors lead to the conclusion that approximatively 75 \%
of the sample analyzed by  Cowley \& Hartwick (1991) has formed
less than 2 Gyr ago, during a recent stellar burst. The velocity
dispersion of 20 km/s refers therefore to young objects whereas
the bulk of the \lmc population is not dominated by young stars
(Geha et al. 1998) and could be associated to much larger velocity
dispersions. As a matter of fact, Hughes et al. (1991) have
measured the radial velocities of long period variables (LPV) in
the \lmc and showed that while young and intermediate LPV's had a
dispersion velocity in the range between 12 and 25 km/s, the old LPV's
had a significantly larger dispersion velocity ranging from 31 to 45 km/s.

The increase of the velocity dispersion with the age of stars is a
well-known fact. Bienaym{\'e} et al. (1987) have modeled stellar populations
in the Milky Way disk and showed that while $\sigma_{W} \sim 6$ km/s at birth,
it increases up to 25 km/s for an age of $7-10$ Gyr. That is why the
characteristic thickness for O and B stars which formed very recently is
$\sim$ 200 pc whereas it reaches $\sim$ 700 pc for solar type objects.
Wielen (1977) has shown that the velocity dispersion $\sigma$ and the
age $\tau$ of a stellar population are related by
\begin{equation}
\sigma^{2} \; = \; \sigma_{0}^{2} + C_{\sigma} \tau \;\;.
\label{WIELEN}
\end{equation}
where $\sigma_{0}$ is the initial velocity dispersion.
As time goes on, disk stars undergo dynamical shocks with, for
instance, molecular clouds. The diffusion of stellar orbits caused
by such gravitational inhomogeneities results in older populations
being associated with larger velocity dispersions. Stars diffuse in velocity
space. For the
Milky Way, Wielen obtained a global diffusion coefficient
\begin{equation}
C_{\sigma} \simeq 6. \times 10^{-7} \; {\rm km^{2} \, s^{-2} \, yr^{-1}}
\end{equation}
whereas he got a value 6 times smaller for vertical motions.

We estimate therefore that the quoted value $\sigma_{W} \sim$ 20 km/s
is only representative of the youngest stars and that the bulk of
the \lmc populations are older, hence with larger velocity dispersions.
The favored star formation history (model e of Geha et al. 1998) has a
star formation rate which remains constant for 10 Gyr and then 
increases by a factor of three for the past 2 Gyr. The average stellar
age is $\sim$ 5 Gyr, quite similar to the age of the Milky Way disk
populations. For the latter, $\sigma_{W}$ increases from 6
to 25 km/s as stars grow old. Scaling these results to the \lmc
implies that the oldest populations should reach there a dispersion
velocity $\sigma_{W} \sim$ 80 km/s. The resulting optical depth
would be therefore an order of magnitude larger than Gould's estimate,
in fairly good agreement now with the observations.

At this point, a detailed investigation of the effect of an increased
velocity dispersion on the optical depth for \lmc self-lensing is
mandatory. To start with, we have refined in section~2 Gould's
description by introducing several isothermal stellar components.
Depending on the stellar species, the velocity dispersion $\sigma_{W}$
varies from 20 up to, at most, 80 km/s. This corresponds to a
diffusion coefficient $C_{\sigma}$ lying in the range from 0 to $5
\times 10^{-7} \; {\rm km^{2} \, s^{-2} \, yr^{-1}}$. The
gravitational potential satisfies a Poisson equation. From the
determination of the vertical distribution of stars, the self-lensing
optical depth and the event rate are calculated, along with the
distribution of event durations.  For our extreme model, the optical
depth is $\sim 0.7-1 \times 10^{-7}$ in the inner $3^{\circ}$. As a
first approach, the variation of the surface mass density $\Sigma$
with radius $R$ is assumed to follow a Mestel profile (see section
\ref{sec:disk}). We have also investigated the tidal effects that our
Galaxy generates on the stellar populations of the \lmc.
Notice that in this thin-disk model, the vertical and radial directions are
somewhat disconnected. At radius $R$, the stellar distribution is derived as
if the \lmc disk looked infinite. We have therefore considered in
section~3 the case of a multi-component model where each stellar
population is distributed according to an oblate spheroid, with significant
flattening along the disk. The velocity dispersion $\sigma_{W}$ and the
flattening depend on the stellar species. In the extreme model,
$\sigma_{W}^{\rm max}$ reaches 45 km/s, hence a diffusion coefficient $C_{\sigma}
\sim 1.4 \times 10^{-7} \; {\rm km^{2} \, s^{-2} \, yr^{-1}}$,
not too far from the Milky Way value. Finally, in section~4,
we show that adopting a larger value for the \lmc rotation velocity 
results into an enhanced self-lensing rate. We suggest further strategies to
test our models and we conclude in section~\ref{sec_conclusion}.

\section{The Large Magellanic Cloud as a Thin Mestel Disk}
\label{sec:disk}
\subsection{Description of the model} 
\label{ssec:descript_dsk}

The bulk of the
mass of the \lmc lies in a nearly face-on disk, with an inclination of 
27$^{\circ}$. The Cloud exhibit a bar structure emitting about 10 to 15~\%
of the total luminosity, but which is not included in the models presented here.
This disk extends at least to $8^{\circ}$ from the center but may
conceivably be larger. The mass in the inner $\sim 4-5$ kpc
($1^{\circ}$ = 0.92 kpc for a distance modulus of 18.6) is estimated
by Hughes et al. (1991) to be
$\leq 6.2 \pm 1.5 \times 10^{9} \; \Msol$ from the 
dispersion of the radial velocities of Long Period Variables (LPV) stars.
The virial mass is smaller: $4.4 \times 10^{9} \; \Msol$.

Measurements of the velocity dispersion of planetary nebulae and star
clusters up to $\sim 8^{\circ}$ radius as well as HI observations indicate
that the \lmc disk rotates. The rotation curve appears flat out to about
$8^{\circ}$ radius, and may remain constant out to $15^{\circ}$. As 
the rotation curve of the \lmc does not show signs of a Keplerian 
falloff, the mass in the inner $8^{\circ}$ radius is inferred
(Schommer et al. 1992) to be $\leq 1.5 \pm 2.5 \times 10^{10} \; \Msol$.
Because the \lmc disk is nearly face-on, the rotation velocity is hard
to determine accurately. It lies in the range between 50 and 80 km/s.
From now on, we adopt the median value of $V_{C} = 65$ km/s, slightly
smaller than the value of 77 km/s favored by Schommer et al. (1992).
Our \lmc mass is therefore on the low side. Note that our approach
is meant to be conservative. Should the \lmc be more massive, our
conclusions would be strengthened. As shown in section~4, a larger
rotation velocity would actually result into an enhanced rate for
gravitational self-lensing.

The surface mass density at radius $R$ follows a Mestel profile
\begin{equation}
\Sigma(R) \; = \; 156 \, [\Msol \, {\rm  pc}^{-2}] \;
\left( \frac{V_{C}}{65 \, {\rm km \, s^{-1}}} \right)^{2} \;
\left( \frac{1 \, {\rm kpc}}{R} \right) \;\;.
\label{SIGMA_MESTEL}
\end{equation}
This implies a \lmc mass of
\begin{equation}
M(R) \; = \; 4.9 \times 10^{9} \, \Msol \;
\left( \frac{V_{C}}{65 \, {\rm km \, s^{-1}}} \right)^{2} \;
\left( \frac{R}{5 \, {\rm kpc}} \right) \;\;.
\end{equation}
Increasing the rotation velocity from 65 to 80 km/s results in an 
increase of the mass in the inner 5 kpc from $4.9$ to $7.4 \times 
10^{9} \, \Msol$.  Both values are in fair agreement with the value 
$4.7-7.7 \times 10^{9} \, \Msol$ of Hughes et al.  (1991).

We have modeled the disk to contain several stellar populations.
We assumed that a Salpeter distribution holds for the surface mass
densities:
\begin{equation}
\frac{d\Sigma}{dm} \propto m^{-\alpha} \;\; ,  \label{eq:Salpeter}
\end{equation}
where $\alpha = 1.35$. Above 0.5 $\Msol$, this is consistent
with HST stellar counts in the \lmc as discussed by
Geha et al. (1998). The normalization factor is obtained
from the requirement that the global surface mass density
$\Sigma(R)$ is recovered once the contributions $d\Sigma$
of the various stellar components are added together.
The luminosity of a \ms star is related to its mass
through $L [\Lsol] = (m [\Msol])^\beta$, with
$\beta \approx 4$ (Henry \& McCarthy 1993). The stars of the
\lmc that are monitored by the microlensing experiments \eros
and \macho have an absolute magnitude smaller than $M_{V} \sim 1$.
This corresponds to stellar masses $\sim 2.2-2.4 \; \Msol$.
We have therefore extended our stellar distribution up to
$m_{\rm max} = 2 \, \Msol$. In our simple model, this massive component
mimics the brightest members of the \lmc populations, such as
the CH stars. It also stands for the sources which may be
potentially amplified by other \lmc members. Notice that our results
are not sensitive to the exact value of $m_{\rm max}$. The lightest stars of
our model have a mass $m_{\rm min} = 0.1 \, \Msol$. They correspond to the
faintest objects on the \ms, the red-dwarves. The mass-to-light ratio
of our stellar distribution is readily derived
\begin{equation}
\frac{M}{L} \; = \; \frac{\beta - \alpha}{\alpha -1} \times 
{\displaystyle \frac
{\displaystyle m_{\rm min}^{1-\alpha} - m_{\rm max}^{1-\alpha}}
{\displaystyle m_{\rm max}^{\beta-\alpha} - m_{\rm min}^{\beta-\alpha}}} \;\; .
\end{equation}
With the values $\alpha = 1.35$ and $\beta = 4$, this implies
$M/L = 1.75$ in solar units.
At a radius of 1 kpc, the surface brightness is
$\sim 90 \, \Lsol \, {\rm pc}^{-2}$. Relation \ref{SIGMA_MESTEL}
gives a surface mass density of $156 \, \Msol \, {\rm pc}^{-2}$,
hence a mass-to-light ratio of 1.71, in good agreement with
our model (the effect of extinction is neglected). Note 
that in the case of the oblate spheroids of
section~3, the surface mass density
$\Sigma(1 \, {\rm kpc}) \sim 220 \Msol \, {\rm pc}^{-2}$ and
the mass-to-light ratio becomes 2.4. Those values may be compared
to the mass-to-light ratio of our galactic disk where $M/L \sim 2$.
As regards the \lmc, the general trend is an increase of $M/L$ with
the radius $R$. At 4 and 5 kpc, the thin Mestel disk model respectively
gives $M/L$ = 4.3 and 5.3 whereas, for the flattened spheroids,
we get 5 and 6. This is in agreement with Schommer et al. (1992)
who conclude that the integrated mass-to-light ratio of the \lmc
is $\sim$ 10. That increase of $M/L$ with $R$ might be due to an
evolution of the stellar composition along the disk. The farther
away from the center a stellar field lies, the fewer bright stars
it contains. This should lead to different HR diagrams at different
locations in the \lmc disk. For simplicity, we have not dealt
here with such refinements and we have kept our stellar distributions
independent of the position $R$. We leave to section~4 the discussion
of the influence of $M/L$ on our results.

The vertical velocity dispersion $\sigma_{W}$ depends on the
stellar population. The only relevant dynamical parameter is
actually the relationship between the surface mass density
and $\sigma_{W}$. Stars of different masses may well have the
same velocity dispersions. If so, their vertical distributions
are similar, with identical disk thickness. Relation
\ref{WIELEN} implies that stars with the same age should have
the same velocity dispersion. As the \lmc formed $\sim$ 12 Gyr
ago, stars whose mass $m < 1 \, \Msol$ have not yet left the \ms.
They have the same average age of 6 Gyr. Their velocity
dispersion had ample time to diffuse and increase. That is why our
first stellar bin comprises stars with masses between 0.1 $\Msol$
(red-dwarves) and 1 $\Msol$ (solar type objects). Their velocity
dispersion $\sigma_{W}^{\rm max}$ is the largest of all species.
Between 1 and 2 $\Msol$ (our most massive bin), the dispersion
$\sigma_{W}$ is obtained from the diffusion equation (\ref{WIELEN})
and from noticing that the \ms lifetime decreases as $m^{-3}$.
This leads to the relation
\begin{equation}
\sigma_{W}^2 \; \simeq \; A \, + \, \frac{B}{m^{3}} \;\; .
\label{sigma_mass_relation}
\end{equation}
The coefficients $A$ and $B$ are determined by requiring that
the dispersion velocity increases from $\sigma_{0} = 20$ km/s
for the brightest and heaviest objects up to
$\sigma_{W}^{\rm max}$ for the light stars.
Table~1 features the velocity dispersions of the various stellar
components in the case of model~A where the radius $R$ = 1 kpc
and $\sigma_{W}^{\rm max} = 80$ km/s. In the sequel, we will assume 
an horizontal velocity dispersion equal to the vertical velocity
dispersion.

\begin{table}
\caption[]{Model A corresponds to the radius $R$ = 1 kpc,
a rotation velocity of 65 km/s and
$\sigma_{W}^{\rm max} = 80$ km/s.
}
\label{model_A}
\[
\begin{array}{|c|c|c|c|} \hline \hline
m [\Msol] &
\sigma_{W} \mbox{[km/s]} &
10^{4} \times \rho_{c} [\Msol \mbox{pc}^{-3}] &
\Sigma [\Msol \mbox{pc}^{-2}] \\ \hline \hline
0.1-1 & 80   & 226    & 134.4 \\
1.1      & 68.5 &   6.73 &   3.3 \\
1.2      & 59.2 &   7.13 &   2.9 \\
1.3      & 51.6 &   7.52 &   2.6 \\
1.4      & 45.2 &   7.91 &   2.4 \\
1.5      & 39.7 &   8.33 &   2.2 \\
1.6      & 34.9 &   8.78 &   2.0 \\
1.7      & 30.6 &   9.31 &   1.8 \\
1.8      & 26.8 &   9.93 &   1.7 \\
1.9      & 23.3 &  10.7  &   1.6 \\
2        & 20   &  11.7  &   1.5 \\ \hline \hline
\end{array} \]
\end{table}

Each model is specified by the radius $R$ that has been varied
from 1 to 5 kpc. The regions surveyed by \eros and \macho are actually
in the inner $5^{\circ}$ from the \lmc center. The value
of $\sigma_{W}^{\rm max}$ has been varied from 20 to 80 km/s
and the rotation velocity has been set equal to $V_{C} = 65$ km/s.
Once a model is specified, the distribution of surface mass
densities and vertical velocity dispersions ensues from
the arguments discussed above. Each stellar component $i$ is
assumed to be isothermal, with vertical velocity dispersion
$\sigma_{W,i}$. It is distributed according to
\begin{equation}
\rho_{i}(z) \; = \; \rho_{i,c} \;
\exp \left\{ - \frac{\Phi(z)}{\sigma_{W,i}^{2}} \right\} \;\; .
\end{equation}
We have solved the Poisson equation along the vertical direction $z$
\begin{equation}
{\displaystyle \frac{d^{2} \Phi}{dz^{2}}} \; = \; 4 \pi G \, \rho(z) \;\; ,
\end{equation}
where the total mass density $\rho(z)$ comprises the various
stellar species. The gravitational potential $\Phi(z)$ and
its derivative $\Phi'(z)$ are set equal to 0 at the origin.
The Poisson equation is integrated up to $z_{\rm max} = \pm 10$ kpc.
Stars located beyond that limit cannot really be considered
as members of the \lmc system. For each stellar population $i$,
we compute the surface mass density $\Sigma_{i}$ for
$|z| \leq 10 \, {\rm kpc}$ and modify its
central mass density $\rho_{i,c}$ in such a way that, after a
dozen iterations, the proper distribution of surface mass
densities is recovered. The convergence is found to be excellent.

Should the \lmc disk contain a single stellar species and
extend vertically to infinity, we would have obtained a
surface mass density of
\begin{equation}
\Sigma = 2 \sqrt{2} \rho_{c} a \;\; ,
\end{equation}
where the typical thickness $a$ is defined as
\begin{equation}
a \; = \; \frac{\sigma_{W}}{\sqrt{4 \pi G \rho_{c}}} \;\; .
\end{equation}
Assuming a Mestel profile to accommodate the flat rotation
curve of the \lmc would have led to the conclusion that the
thickness at radius $R$ is
\begin{eqnarray}
a & = & 1.34 \times 10^{2} \, {\rm pc}
\left( \frac{\sigma_{W}}{20 \, {\rm km/s}} \right)^{2}
\left( \frac{V_{C}}{65 \, {\rm km/s}} \right)^{-2} \nonumber \\
& \times &
\left( \frac{R}{1 \, {\rm kpc}} \right) \;\; .
\end{eqnarray}
This simple picture is complicated by the presence of several
stellar populations with different velocity dispersions
and also by the presence of a cut-off at $\pm$ 10 kpc.
As is clear from table~\ref{model_A}, the low-mass stars
with $m = 0.1 - 1 \, \Msol$ dominate the disk dynamics.
The optical depth $\tauLMC$ for self-lensing ensues
from relation~(\ref{GOULD}) where $\sigma_{W}^{\rm max}$ stands
for the velocity dispersion.
However, Gould's limit is only recovered in the \lmc regions where the 
disk thickness $a$ is much smaller than the vertical boundary $z_{\rm 
max} = \pm 10$ kpc.  This is the case for $R = 1$ kpc and for 
$\sigma_{W}^{\rm max} = 20$ km/s where $a = 130$ pc.  When the 
velocity dispersion of low-mass stars is increased up to 80 km/s or 
when the radius $R$ is large, the \lmc disk is no longer thin with 
respect to the boundary $z_{\rm max}$.  The actual value of $\tauLMC$ 
is smaller than what would naively be inferred from a mere scaling 
from an infinite disk. This will be further discussed in section~4.

\subsection{Microlensing parameters}
\label{ssec:miclens_dsk}

For each of the models, one can compute the total optical depth $\tau$ 
and the event rate $\Gamma$.  The contribution of sources of mass $m_s$ located at 
$z_s$ and deflectors of mass $m_d$ to these microlensing parameters is 
given by:
\begin{multline}
  \delta\tau(m_d, z_s) =  \\ 
    \int_{-z_{\rm max}}^{z_s} \!\! dz_d \,
    \frac{\rho_d(z_d \cos\theta,m_{d})}{m_d}\, \pi R_E(m_d,z_d,z_s)^2 
\label{eq_dtau1}
\end{multline}
\begin{multline}
  \delta\Gamma(m_d, m_s, z_s) =  \\ 
    \int_{-z_{\rm max}}^{z_s} \!\! dz_d \,
    \frac{\rho_d(z_d \cos\theta,m_{d})}{m_d} \, 2 R_E(m_d,z_d,z_s) 
    <\!\!\left|v_{\rm rel}\right|\!\!>
\label{eq_dgamma1}
\end{multline}
where $\theta$ is the inclination angle of the \lmc (taken equal to
27 degrees), $z$ is measured along the line-of-sight, and $\rho_d$ 
is the mass density of the deflectors. 

The velocities $v_s$ and $v_d$ of the objects are supposed to be
Gaussian-distributed with a mass-dependent velocity dispersion, as
explained in section \ref{ssec:descript_dsk}.  The term
$<\!|v_{\rm rel}|\!>$ in equation (\ref{eq_dgamma1}) represents the average
of the relative velocity between the deflector and the source over all
possible velocities of these two objects:
\begin{equation}
<\!|v_{\rm rel}|\!> = \!\int\!\!\!\int\! f_s(v_s,m_{s}) \, f_d(v_d,m_{d}) \, |v_s - v_d|\, dv_s\, dv_d
\end{equation}
where $f_s$ (resp. $f_d$) is the source (resp. deflector) velocity distribution
function --- projection effects are neglected since $z \ll \DLMC$.

Let $\rho_s$ be the mass density of the sources. The mass density at a
given coordinate only depends on the mass of the object. 
Integrating equations (\ref{eq_dtau1}) and (\ref{eq_dgamma1}) over
all source positions yields:
\begin{equation}
\tau(m_d, m_s) =
  \frac{\int_{-z_{\rm max}}^{z_{\rm max}} dz_s
    \frac{\rho_s(z_s \cos\theta, m_{s})}{m_s} \, \delta\tau(m_d, z_s)
       }
       {\int_{-z_{\rm max}}^{z_{\rm max}} dz_s \;\frac{\rho_s(z_s \cos\theta,m_{s})}{m_s} 
       }\label{eq_dtau}
\end{equation}
\begin{equation}
\Gamma(m_d, m_s) =
  \frac{\int_{-z_{\rm max}}^{z_{\rm max}} dz_s
    \frac{\rho_s(z_s \cos\theta,m_{s})}{m_s} \;\delta\Gamma(m_d, z_s)
       }
       {\int_{-z_{\rm max}}^{z_{\rm max}} dz_s \; 
       \frac{\rho_s(z_s \cos\theta,m_{s})}{m_s} 
       } \label{eq_dgamma}
\end{equation}

The values obtained with the equations (\ref{eq_dtau}) and
(\ref{eq_dgamma}) must then be summed on all the deflector species, and
averaged on all the source species. Only the uppermost mass bin will be
considered as possible sources (i.e. $m_s \in [1.9--2.0]\, M_\odot$)
in agreement with the mass of the faintest stars that are
reconstructed in microlensing surveys, whereas all the mass bins are
of course taken into account when summing for all possible
deflectors. The global optical depth and event rate due to the entire
stellar population are therefore given by:
\begin{align}
\tau &= \sum_{i_d=1}^{i_d=i_{\rm max}} \tau(m_d(i_d),m_s(i_{\rm max})) \\
\Gamma &= \sum_{i_d=1}^{i_d=i_{\rm max}} \Gamma(m_d(i_d),m_s(i_{\rm max}))
\end{align}
The first mass bin has in practice to be treated specially,
since it is not thin. The deflector masses in this bin follow the
Salpeter law given in Eq.~(\ref{eq:Salpeter}).

\begin{figure}[h] \begin{center} 
  \epsfig{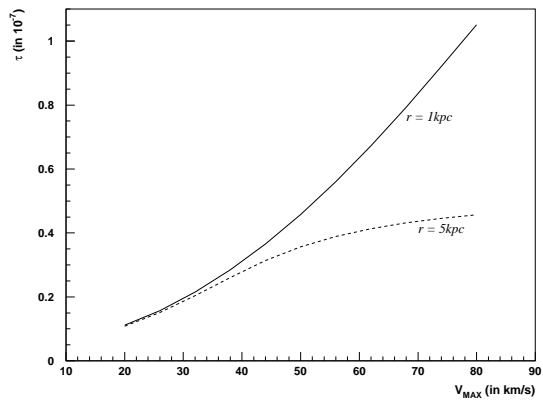} \caption{Evolution of the optical
  depth $\tau$ with the maximum velocity dispersion of the various
  populations for the disk model, at two different distances from the
  center of the \lmc: 1~kpc and 5~kpc.}  \label{fig1}\end{center}
\end{figure}

\begin{figure}[h] \begin{center} 
  \epsfig{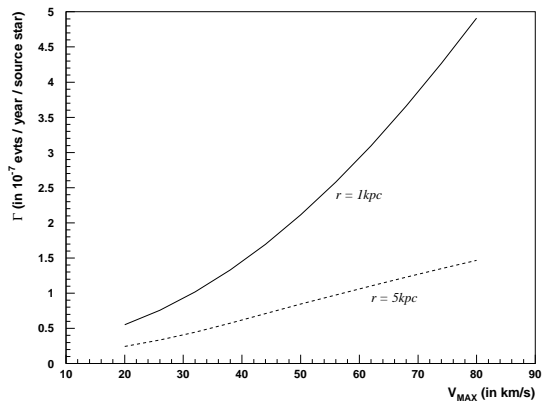} \caption{Evolution of the event
  rate $\Gamma$ with the maximum velocity dispersion of the various
  populations for the disk model, at two different distances from the
  center of the \lmc: 1~kpc and 5~kpc.}  \label{fig2}\end{center}
\end{figure}

These results depend on the maximum dispersion velocity of the stars
(i.e. on the dispersion velocity of the lightest population), as
illustrated in figures \ref{fig1} and \ref{fig2}.
The configuration where all the stellar populations have the same
velocity dispersion (i.e. when $v_{\rm max} = 20$~km/s) yields the
same results as the calculations done by Gould (Gould 1995).  The
slower increase in the optical depth with the maximum velocity for a
radial distance of 5~kpc as compared to that observed at a distance of
1~kpc is due to the 10~kpc vertical cut-off we have imposed on the
stars to be considered as \lmc objects (see comment at the end of
section \ref{ssec:descript_dsk}), since the disk gets wider on the
edges. As expected, the optical depth and event rate decrease as one
goes to larger and larger radii from the galactic center. For $v_{\rm
max} = 80$~km/s, there is a factor of two in optical depths and three
in event rates between a distance of 1 and 5~kpc.

\begin{figure}[h] \begin{center} 
  \epsfig{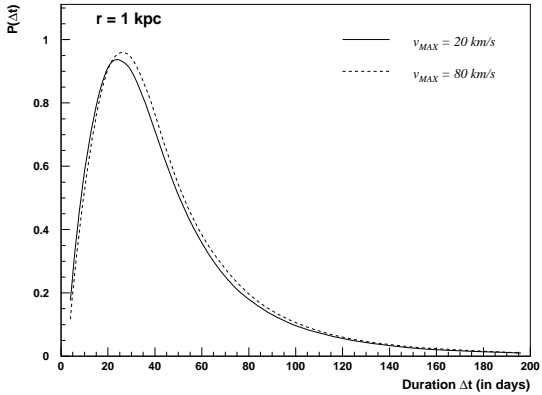} \caption{Predicted distribution
  of event durations for the multi-mass disk model at a distance of
  1~kpc from the center of the \lmc. Plain curve is for a maximum
  velocity dispersion of 80~km/s, dashed curve for a maximum velocity
  dispersion of 20~km/s.}  \label{fig3a}\end{center}
\end{figure}

\begin{figure}[h] \begin{center} 
  \epsfig{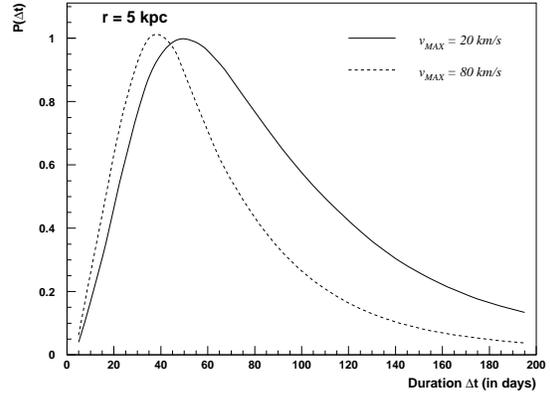} \caption{Predicted distribution
  of event durations for the multi-mass disk model at a distance of
  5~kpc from the center of the \lmc. Plain curve is for a maximum
  velocity dispersion of 80~km/s, dashed curve for a maximum velocity
  dispersion of 20~km/s.}  \label{fig3b}\end{center}
\end{figure}

An interesting comparison to the microlensing data can be done through 
the predicted distribution of event durations, $P(\Delta t)$ which can 
be calculated directly from the event rate: $P(\Delta t) \propto 
d\Gamma/d\Delta t$.  In a similar manner as what was done for the 
calculation of the optical depth and the event rate, we have:
\begin{equation}
\begin{split}
    \frac{d\Gamma}{d\Delta t}(\Delta t, m_d, m_s, &z_s) =   \\
    \int_{-z_{\rm max}}^{z_s} \!\!\!  &dz_d 
     \frac{\rho_d(z_d \cos\theta,m_{d})}{m_d} \times \\
    \,\int\!\!\!\int\   &dv_{d} \, dv_{s} \, 
     f_{d}(v_{d},m_{d}) \,   f_{s}(v_{s},m_{s}) \times \\
    & \delta\left(\frac{R_E}{|v_s - v_d|}-\Delta t\right)
    \, 2 R_E |v_s - v_d|  
\end{split}
\end{equation}
Integrating over the source positions yields:
\begin{equation}
\begin{split}
  \frac{d\Gamma}{d\Delta t}&(\Delta t,m_d,m_s) =  \\
  &\frac{\int_{-z_{\rm max}}^{z_{\rm max}} dz_s \,
    \frac{\rho_s(z_s \cos\theta,m_{s})}{m_s} \;\frac{d\Gamma}{d\Delta t}(\Delta t,m_d,m_s,z_s)
       }
       {\int_{-z_{\rm max}}^{z_{\rm max}} dz_s \; \frac{\rho_s(z_s \cos\theta,m_{s})}{m_s} 
       } \label{eq_ptau}
\end{split}
\end{equation}
and therefore
\begin{equation}
\frac{d\Gamma}{d\Delta t}(\Delta t) = \sum_{i_d=1}^{i_d=i_{\rm max}} 
\frac{d\Gamma}{d\Delta t}(\Delta t,m_d(i_d),m_s(i_{\rm max}))
\end{equation}
Figures \ref{fig3a} and \ref{fig3b} show the distribution of the event
duration obtained in this multi-mass disk model of the \lmc
normalized to a maximum value of 1, for two different values of the
maximum velocity dispersion and at two different distances from the
galactic center.

At 1~kpc from the center, the duration distribution is nearly 
independent of the maximum velocity dispersion, because the velocity
cancels out in $R_{E}/v$ when $z \propto v^{2}$. At 5~kpc from the
center, the truncation at $z = 10\; {\rm kpc}$ prevents this 
cancellation.

\subsection{Tidal effects}
We have finally modeled the tidal effects which the Milky
Way potentially induces on the optical depth $\tauLMC$. Because
the \lmc rotates in the potential well of our Galaxy, the interplay
between the gravitational attraction and the centrifugal force
results into an effective tidal potential
\begin{equation}
\Phi_{\rm tide} (z) \simeq - \frac{3}{2} \,
\frac{G M_{\rm gal}}{\DLMC^{3}}\;\cos^{2}i \; z^{2} \;\; ,
\end{equation}  
in the vicinity of the disk, for small $z$. The associated
repulsion tends to stretch the stellar populations apart
from the disk as if they were no longer pressure supported.
The Poisson equation now becomes
\begin{equation}
{\displaystyle \frac{d^{2} \Phi}{dz^{2}}} \; = \; 4 \pi G \,
\left\{ \rho(z) - \bar{\rho}_{\rm gal} \right\} \;\; ,
\end{equation}
where $\bar{\rho}_{\rm gal}$ denotes the average density of
our Galaxy as seen by the \lmc at a distance of
$\DLMC$ = 52 kpc. With a galactic mass of
$M_{\rm gal} = 6 \times 10^{11} \, \Msol$, we get
$\bar{\rho}_{\rm gal} \simeq 10^{-3} \, \Msol \, {\rm pc}^{-3}$.
The tidal forces make the potential well of the  \lmc  disk flatter.

\begin{figure}[h] 
 \begin{center} 
  \epsfig{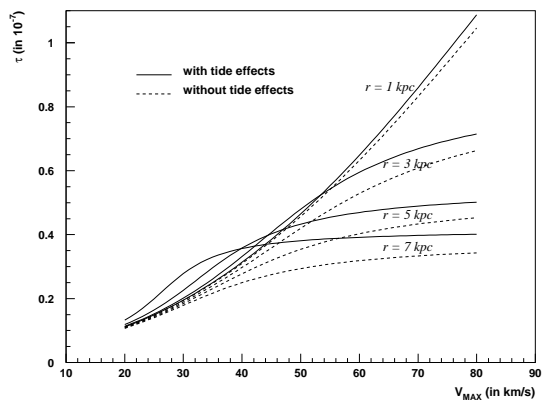} 
  \caption{Optical depth $\tauLMC$ as a function
of the velocity dispersion $\sigma_{W}^{\rm max}$, for various radii
$R$, with and without the inclusion of tidal effects in the model.}  
  \label{fig4}
 \end{center}
\end{figure}

In figure \ref{fig4}, the optical depth $\tauLMC$ is plotted as a function
of the velocity dispersion $\sigma_{W}^{\rm max}$, for various radii
$R$. The trend is an increase of $\tauLMC$ when tidal
effects are included. Near the \lmc center, the disk is thin with
a strong cohesion. Tidal effects are negligible. Far from
the \lmc center, the vertical scale length of the disk increases,
translating into an enhanced sensitivity of the stellar populations
to the Milky Way tide. The increase of the optical depth is noticeable.
Finally, for $R \geq 5$ kpc, the disk scale length $a$ compares with
the cut-off $z_{\rm max} = \pm 10$ kpc as the dispersion velocity
$\sigma_{W}^{\rm max}$ increases towards 80 km/s. In that case, low-mass
stars tend to be uniformly distributed, even in the absence of tidal
forces.

\section{Ellipsoidal models}
\label{sec:ellipsoid}

\subsection{Description of the model}
\label{sec_descript_ellipso}

In the previous disk model, the $z$ and $R$ dependences of the
potential were decoupled, which is not true for the exact potential of
the \lmc.  In this section, we therefore investigate another model, 
more realistic as far as the co-dependence on $z$ and $R$ is
concerned.  

\begin{figure}[h] 
 \begin{center} 
  \epsfig{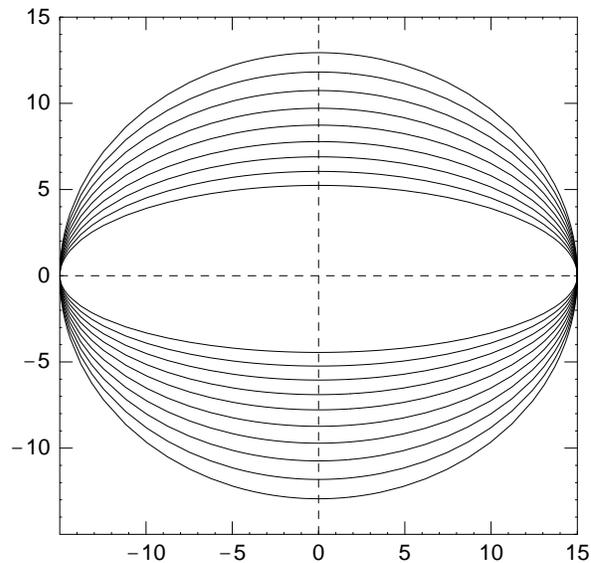} 
  \caption{Truncation ellipsoids, each corresponding to a different
    species. 
Outer ellipsoid corresponds to the 0.1 $\Msol$ - 1 $\Msol$ mass bin, and 
then inwards to the species from 1.1 to 2~$\Msol$, by 0.1 $\Msol$ increments.}
  \label{fig5}
 \end{center}
\end{figure}

The \lmc is described with the same mass bins as in the previous
section. The density of each population is assumed to be of the form
\begin{equation}
\rho_i(R,z) = \frac{\Lambda_i}{\displaystyle R^2 + \frac{z^2}{1-e_i^2}}
\end{equation}
so that the iso-density surfaces are oblate ellipsoids with
ellipticity $e_i$.
This distribution is truncated at the ellipsoid passing through the
cut-off radius $R_{\rm max} = 15\, \mbox{kpc}$, beyond which the mass density
is zero. Note that the cut-off ellipsoids are not the same for the
different mass species. They are illustrated in figure~\ref{fig5}.

The corresponding rotation curve is flat, with a circular velocity
\begin{equation}
v_{\rm rot}^2 (R) = 4 \pi G \sum_i \Lambda_i \sqrt{1-e_i^2} \times
\frac{\arcsin e_i}{e_i}
\end{equation}
The surface mass density of species $i$ is
\begin{equation}
\Sigma_i(R) = \frac{2 \Lambda_i \sqrt{1-e_i^2}}{R}
\arctan \left( \frac{\sqrt{R_{\rm max}^2-R^2}}{R} \right)
\label{surf_dens_ellipsoids}
\end{equation}
With a given set of parameters $(\Lambda_i,e_i)$, the local vertical
velocity dispersion is defined as
\begin{equation}
\sigma_{W,i}^2 = \frac{\Phi(R,z)-\Phi(R,0)}{\log \{
  \rho(R,0)/\rho(R,z) \}}
\end{equation}
We define $v_{W,i}$ as the vertical velocity dispersions
averaged along $z$, which are more closely related to observable quantities.
\begin{equation}
v_{W,i}(R) = \frac{\int_{-\infty}^{\infty} \rho_i(R,z)
  \sigma_{W,i}(R,z) dz}{\int_{-\infty}^{\infty} \rho_i(R,z) dz}
\end{equation}

We computed the parameters $\Lambda_i$ and $e_i$ that realize the
following properties:
\begin{itemize}
\item[$\bullet$] The circular velocity is 65 km/s.
\item[$\bullet$] The surface density mass function is given by a
Salpeter law (see section~2). As can be seen in
(\ref{surf_dens_ellipsoids}), this condition can be consistently
fulfilled for every radius.
\item[$\bullet$] as in section~2, the average vertical velocity
  dispersions $v_{W,i}$ vary according to the mass $m_i$ as given in
  (\ref{sigma_mass_relation}), from $v_{\rm min}$ to $v_{\rm max}$. 
\end{itemize}

Unlike the disk model, there is an upper limit to the
vertical velocity dispersion we can impose.
For a one-species model, this limit would be $\sigma_W^{\rm max} =
V_c/\sqrt{2}$, numerically $46 \, \mbox{km/s}$, given the chosen 
rotation velocity.
This limit is slightly different for a multi-mass model, but still present.
Accordingly, we impose velocity dispersions ranging from $v_{\rm min}=20
\, \mbox{km/s}$ for the heavy stars to $v_{\rm max}=45 \, \mbox{km/s}$
for the light stars. This is model B-1.
For comparison, we also present the results for model B-0 in
which all species have a velocity dispersion of 20 km/s.
The parameters for model B-1 are shown in table~\ref{model_B_1}.

\begin{table}[h]
\caption{Parameters obtained after convergence for model B-1,
i.e. with $v_{\rm max} = 45$~km/s.}
\label{model_B_1}
\[
\begin{array}{|c|c|c|c|} \hline \hline
m [\Msol] & e_i& 10^{3} \times  \Lambda_i [\Msol \mbox{pc}^{-1}] &
\mbox{v}_{W,i} \mbox{[km/s]}\\
\hline \hline
0.1-1 & 0.438 & 70.5 & 45   \\
1.1   & 0.663 & 2.08 & 39.5 \\
1.2   & 0.768 & 2.16 & 35.2 \\
1.3   & 0.830 & 2.22 & 31.8 \\
1.4   & 0.869 & 2.27 & 29.1 \\
1.5   & 0.895 & 2.29 & 26.8 \\
1.6   & 0.913 & 2.30 & 24.9 \\
1.7   & 0.927 & 2.30 & 23.4 \\
1.8   & 0.937 & 2.29 & 22.1 \\
1.9   & 0.945 & 2.26 & 20.9 \\
2     & 0.951 & 2.23 & 20 \\ \hline \hline
\end{array} \]
\end{table}

\subsection{Microlensing parameters}

\begin{figure}[h] \begin{center} 
  \epsfig{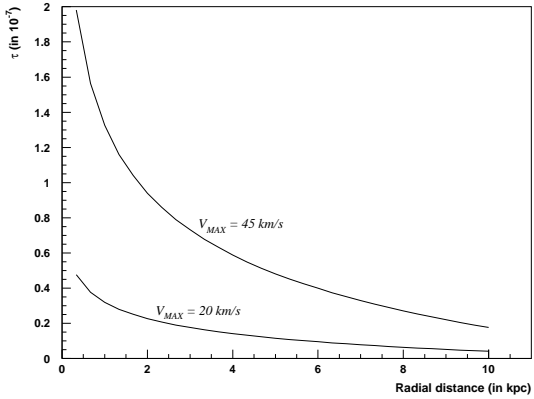} \caption{Evolution of the optical
  depth $\tau$ with the distance to the center of the \lmc, for the
  ellipsoidal model and two values of the maximum velocity dispersion:
  20 km/s (model B-0) and 45 km/s (model B-1).}  \label{fig6}\end{center}
\end{figure}

\begin{figure}[h] \begin{center} 
  \epsfig{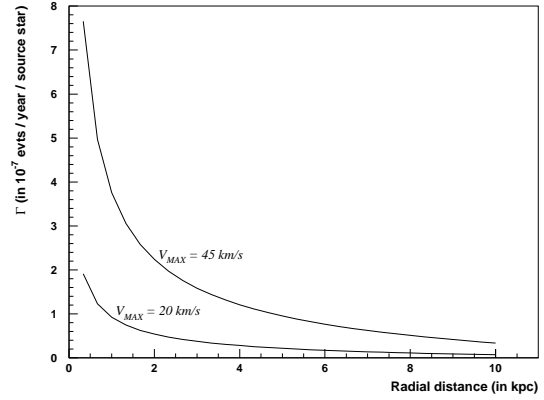} \caption{Evolution of the event
  rate $\Gamma$ with the distance to the center of the \lmc, for the
  ellipsoidal model and two values of the maximum velocity dispersion:
  20 km/s (model B-0) and 45 km/s (model B-1).}  \label{fig7}\end{center}
\end{figure}

In a similar way as that explained in section \ref{ssec:miclens_dsk},
it is possible to compute the optical depth and event rate for an
ellipsoidal model of the \lmc. They are shown in figures \ref{fig6}
and \ref{fig7} as a function of the distance to the center of the
\lmc.
There is a strong decrease of both quantities with the radius, which 
can be used to test statistically the location of the lenses detected 
in microlensing experiments.  Note that the effect is even more 
drastic in the ellipsoidal model than in the disk model of the \lmc: 
in the latter, the increased thickness at large radius was 
compensating the decrease in surface density.  

\begin{figure}[h] \begin{center} 
  \epsfig{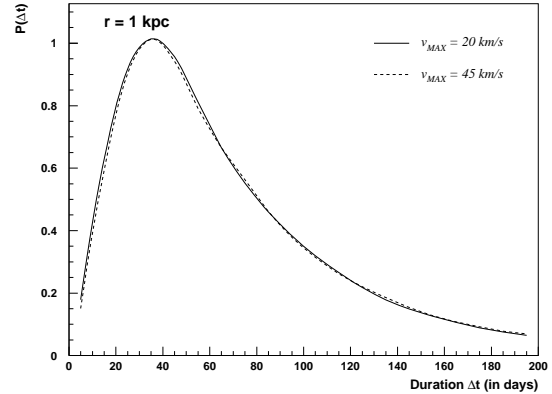} \caption{Predicted distribution
  of event durations for the multi-mass ellipsoidal model at a
  distance of 1~kpc from the center of the \lmc. Plain curve is for 
  model B-1, dashed curve for model B-0.}  \label{fig8}\end{center}
\end{figure}

\begin{figure}[h] \begin{center} 
  \epsfig{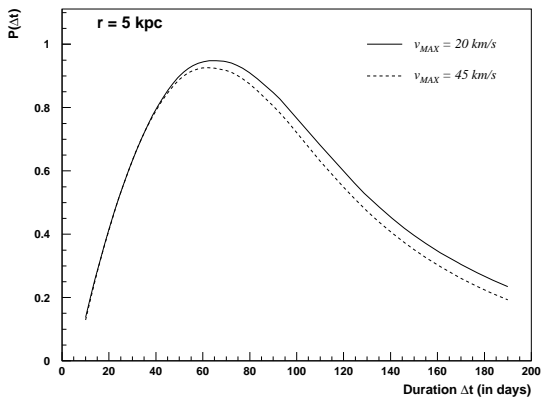} \caption{Predicted distribution
  of event durations for the multi-mass ellipsoidal model at a
  distance of 5~kpc from the center of the \lmc. Plain curve is for 
  model B-1, dashed curve for model B-0.}  \label{fig9}\end{center}
\end{figure}

The predicted distributions of event durations, normalized to a
maximum value of 1, are shown in figures \ref{fig8} and \ref{fig9} at
two different distances from the center of the \lmc.
Since there 
is no cut-off in the ellipsoidal model, the duration distribution is 
nearly independent of the maximum velocity dispersion at all radii.


\section{Discussion and conclusion} 
\label{sec_conclusion}

We have presented here two models for the \lmc, both consistent with
the known properties of the Cloud (rotation velocity, surface light
density), allowing for a multi-mass stellar population, each
associated with a specific velocity dispersion. This correlation
between mass and velocity dispersion is justified by the fact
that lighter stars are on the average older and had the time to
diffuse in velocity space. 


These models of the \lmc predict an optical depth in agreement with
that observed by the \macho and the \eros collaborations: at a
distance of 1~kpc from the \lmc center, the disk model yields
$\tauLMC({\rm disk}) \sim 1\times 10^{-7}$ for a velocity dispersion of the
lightest and therefore oldest objects of 80~km/s, and the ellipsoidal
model yields $\tauLMC({\rm ellipsoids}) \sim 1.3\times 10^{-7}$ for $v_{\rm
max} = 45$~km/s. 

The optical depth which we have derived results from the interplay of 
various effects.  In the case of the disk model, for instance, Gould's 
estimate is only recovered in the thin disk limit.  Setting the 
various stellar velocities at the same value of 80~km/s and pushing 
the \lmc at infinite distance from the Milky Way, we get $\tauLMC = 
1.80 \times 10^{-7}$, in agreement with relation (\ref{GOULD}).  If 
now we assume the same distribution of stellar velocities as in 
Table~\ref{model_A}, \ie with velocities ranging from 20 to 80~km/s, 
the optical depth becomes $1.14 \times 10^{-7}$.  Because the sources 
have small velocity dispersions, they are fairly concentrated towards 
the disk, hence a decrease of $\tauLMC$ by a factor of $\sim \log 2$.  
The proeminent effect is actually the presence of the cut-off $z_{\rm 
max}= \pm 10\;{\rm kpc}$ which we have enforced.  At 3~kpc from the 
center, the optical depth decreases from $1.14 \times 10^{-7}$ to 
$7.57 \times 10^{-8}$.  At small radii, the surface mass density is 
large and the disk thickness $a$ is small as compared to $z_{\rm 
max}$.  There, the decrease of $\tauLMC$ is less sensitive than 
towards the outer fringes of the \lmc disk, where the cutoff 
suppression is quite noticeable.  Taking furthermore into account the 
finite distance of 52~kpc to the \lmc implies a minor decrease of the 
optical depth which becomes now $\tauLMC(3\;{\rm kpc}) = 6.63 \times 
10^{-8}$.  Finally, tidal forces tend to stretch the disk, hence a 
final value of $\tauLMC(3\;{\rm kpc}) = 7.15 \times 10^{-8}$.

If the rotation velocity of the \lmc $V_{C}$ is set equal to 80~km/s 
instead of 65, the main impact on our disk model is an increased 
surface mass density. As the disk becomes thinner, the suppression 
of the optical depth resulting from the above mentioned cut-off is 
less sensitive, hence an increased optical depth at large radii. At 
3~kpc, the optical depth becomes $9.08 \times 10^{-8}$.

In order to model various M/L ratios, we have simply modified the 
relative contribution of our first stellar bin.  That population 
corresponds to faint stars.  Still with $V_{C}=80\;{\rm km/s}$ and at 
$R =$ 3~kpc, an increase of the M/L ratio by a factor of 2 and 5 respectively 
leads to 9.43 and $9.67 \times 10^{-8}$. The optical depth is thus not 
sensitive to the precise value of the M/L ratio, insofar as the disk 
dynamics is dominated by the faint and low-mass species.

Ellipsoids have a lenticular shape.  As the radius $R$ increases, 
their surface mass density drops, and the system becomes thinner 
towards the edges.  Increasing the \lmc rotation velocity without 
changing the velocity dispersion of the various stellar species does 
not change much the optical depth.  Two opposite effects 
are actually at stake: the first is that the mass densities are larger, 
the second that the stellar populations are flattened towards the 
disk.  If now the dispersion velocities $\sigma_{W}$ are scaled to the 
rotation velocity, in order to maintain the shape of the ellipsoids 
--- which depends to first order on the ratio $\sigma_{W}/V_{C}$ --- 
increasing the rotation speed from 65 to 80~km/s results into a 
significantly larger optical depth.  If we set $\sigma_{W}^{\rm 
max}=55\;{\rm km/s}$ and $V_{C}$ = 80~km/s, the optical depth at 1 and 
3~kpc becomes respectively 2.19 and $1.26 \times 10^{-7}$.  That set 
of values actually correspond to an \lmc mass at 5~kpc of $7 \times 
10^{9} \Msol$ (Hughes et al.  1991) that increases up to $1.4 \times 
10^{10} \Msol$ at 10~kpc (Schommer et al.  1992).

\begin{figure}[h] 
  \begin{center} \epsfig{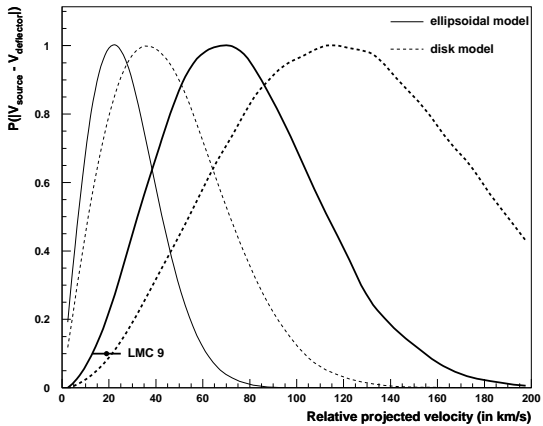}
   \caption{Predicted distribution of relative velocity between source
   and lens, for the ellipsoidal and the disk models, along with the
   measured value for MACHO-LMC-9 (Bennett et al. 1996). Thin curves:
   null horizontal velocity dispersion; thick curves: horizontal
   velocity dispersion equal to the vertical one.} \label{fig10}
   \end{center}
\end{figure}

Reexamining the status of event MACHO-LMC-9 appears worthwhile. As
pointed out in (Bennett, 1996), its measured projected relative
velocity was too low both for a Galactic halo event and for
self-lensing with a standard \lmc disk model.  Figure \ref{fig10}
shows the distribution of the expected projected relative velocity for
the 80 km/s disk and the 45 km/s ellipsoidal models, along with the
measured velocity, for two values of the horizontal velocity
dispersion, null or equal to the vertical one -- the reality being
probably inbetween. The measured value is compatible with an
ellipsoidal model, but only compatible with a disk model if the
horizontal velocity dispersion is not too high.

Reconsidering the shape of the \lmc and the velocity distributions of
its stellar components thus appears to be quite valuable. Two
observational projects could yield interesting results in regards to
this new perspective, and would help to check the validity of the
assumptions made here. Both would be of great consequence for
microlensing experiments.

The first of these is the measure of dispersion velocities for 
RR-Lyrae in the \lmc.  They indeed form a well-defined sample of very 
old stars (more than 10 Gyr old).  If our model is correct, they 
should be found to have a higher velocity dispersion than previously 
measured stars.

The second project is actually a pure extension to the present
microlensing experiments. As evidenced in this paper, self-lensing in
the \lmc can be tested quite accurately with the spatial evolution of
the microlensing parameters such as optical depth and event
rate. Searching for microlensing events on as large an area as
possible over the \lmc would be a very powerful tool to discriminate
between halo and \lmc deflectors, because of the strong decrease of
$\tauLMC$ and $\Gamma_{\mathrm{\mathsc{lmc}}}$ with the distance to
the center of the \lmc in the case of self-lensing. In addition to the
implications such observations would have on the galactic dark matter
issue, they would be of great use for a better understanding of the
structure of the \lmc, the nearest galaxy to the Milky Way.
Furthermore, as shown in (Gould 1998), monitoring as large an area
as possible on the \lmc, using the same exposure time on inner and 
outer fields, optimizes the total number of expected events, even for
Galactic halo lensing. Such a strategy would be even more productive
if associated with a real-time trigger and a high precision follow-up strategy,
yielding high quality microlensing light-curves on which testing for parallax 
effect would provide information on the location of the lenses.

\begin{acknowledgements}
We wish to thank D. Bennett, M.O. Menessier and N. Mowlavi for useful 
discussions, and the members of the EROS collaboration for their comments.
\end{acknowledgements}

\end{document}